# Preprint

# Global Synchronization Protection for Bandwidth Sharing TCP Flows in High-Speed Links

Wolfram Lautenschlaeger
Alcatel-Lucent Bell Laboratories
Stuttgart, Germany
wolfram.lautenschlaeger@alcatel-lucent.com

Andrea Francini
Alcatel-Lucent Bell Laboratories
Mooresville, NC (USA)
andrea.francini@alcatel-lucent.com

*Abstract*— In a congested network link, synchronization effects between bandwidth-sharing TCP flows cause wide queue length oscillations, which may translate into poor link utilization if insufficiently buffered. We introduce global synchronization protection (GSP), a simple extension to the ordinary operation of a tail-drop queue that safely suppresses the flow synchronization. Our minimalistic solution is well suited for scaling with leading-edge link rates: it adds only few extra operations in the fast path and does not require accelerated memory access compared to the line rate. GSP makes it easier to provide advanced control of TCP congestion in high-speed links and in low-power packet processing hardware. Using experiments with a Linux prototype of GSP, we show that, despite its exclusive focus on removing global synchronization, the new scheme performs as well as far more complex active queue management (AQM) schemes like CoDel and PIE.

## I. INTRODUCTION

Packet queues are indispensable in almost all network nodes. They avoid the loss of packets when clustered arrivals temporarily saturate the transmission capacity of a shared link. Typically a queue accumulates few packets, then quickly empties again, but congestion may develop when the saturation of the link capacity becomes persistent. A congested queue grows in size and eventually overflows the buffer space. The resulting loss of packets may degrade the performance of the respective applications. End systems implement congestion control to match the combined traffic offer to the capacity of the most congested link in the data path, so that the occurrence and negative effects of packet losses are minimized.

TCP is today the dominant protocol for congestion control in IP networks. The queues that enable proper operation of TCP are larger than those that resolve temporary contention. With smaller queues TCP still works, but may fail to fill the entire link capacity. To avoid any risk of wasting bandwidth resources, network vendors and operators have been playing safe by scaling buffer sizes with link capacities. As a result, end-to-end data paths include today many network links where large queuing delays can accumulate when congestion occurs. Bloated buffers [1] damage not only interactive applications such as voice/video conferencing and gaming, but also those that require stable throughput, such as adaptive bit-rate (ABR) video streaming.

Suddenly widespread awareness of the bufferbloat issue has created new opportunities to reduce queuing delays everywhere in the network [2]. Combinations of flow queuing (FQ) [3] with active queue management (AQM) schemes for control of the overall buffer occupancy (FQ-CoDel [4] and FQ-PIE [5]) are gaining consensus as the preferred approach for application in home routers and fiber/DSL/cable modems and access nodes [5]. These FQ-AQM schemes hash packet headers onto queues of which the respective flows typically obtain exclusive use. The benefits are flow isolation, fairness, and latency minimization for low-bandwidth, low-delay applications.

The race is far from over in the high-speed links of the network core. The typically large number of concurrent flows in core links discourages the deployment of multi-queue AQM solutions and could in theory make the adoption of small tail-drop buffers a safe choice [6], [7]. However, the same links must ensure high utilization of their capacity also when the number of flows is small and tail-drop can no longer avoid their synchronization. The random early detection (RED) AQM [8], while broadly available today in high-speed routers, is well known for its inability to adapt to varying traffic conditions. Single-queue AQMs of recent introduction (PED [9], CoDel [10], PIE [11]) are certainly more versatile than RED, but their line-rate operation in high-speed links (10 Gb/s and above) is unproven (CoDel in particular may require multiple accesses to the packet memory during a single dequeue operation) and their performance is not always immune from the effects of misconfiguration (throughput losses or delay inflation may occur when the target delay of CoDel and PIE is too small or too large for the round-trip time distribution of the set of active flows).

We introduce Global Synchronization Protection (GSP), a new AQM scheme for high-speed links that reconciles throughput and delay performance with a scalable implementation. Like many pre-existing AQM schemes, GSP achieves the suppression of global synchronization by spreading over time the attribution of packet losses to different flows after congestion builds up a standing queue. The novelty of the scheme versus its predecessors is the simplicity of its operation, which future-proofs it against any foreseeable link rate increase. GSP extends the operation of a conventional tail-drop queue with few fast-path steps that it invokes when it receives a new packet. No extra step is required upon packet departure. Simplicity of operation also implies that the configuration parameters are few, easy to derive from the link capacity and practically insensitive to the traffic conditions.

The paper is organized as follows. In Section II we recall the behavior of a queue loaded with TCP traffic and elaborate on the root cause of global synchronization. In Section III we define the basic GSP algorithm, we illustrate its operating regimes, and select the adaptation strategy for its primary variable. In Section IV we present the results of benchmarking experiments from a 10GbE network of Linux boxes, showing that despite its simplicity GSP is never inferior to any of the single-queue AQM schemes that are most popular today. We draw our conclusions and outline future work in Section V.

## II. GLOBAL SYNCHRONIZATION

Due to the scalability constraint of a single-queue implementation, the primary goal of a buffer management scheme for high-speed links should be the suppression of global synchronization for long-lived TCP flows (loosely defined as flows that remain active long enough to experience a few of the congestion episodes of a bottleneck link). Other nice-to-have features found in FQ-AQM schemes, such as the protection of well-behaved flows from unresponsive ones and the provision of fast lanes to flows of low-bandwidth, low-delay applications, are simply impossible to achieve with a single queue. Still, interactive applications draw important benefits from the buffer size reductions enabled by the suppression of global synchronization. In this section we expose the root causes of global synchronization and their inflating effect on buffer sizes. The discussion is mostly qualitative: we refer to [12] for a detailed quantitative analysis.

### A. Single Flow

In a TCP connection, the transmitter sends data segments over the forward path and receives acknowledgment segments (ACKs) over the reverse path. The ACKs provide confirmation of successful receipt of the data segments by the TCP receiver. The transmitter receives an ACK one round trip time (RTT) after sending the corresponding data segment. The *flight size* is the amount of transmitted data that are yet unacknowledged. The congestion window (*cwnd*) limits the flight size: when *cwnd* is exhausted the transmission of new data can happen only after previously transmitted data are acknowledged [13].

The bit rate of the TCP connection is defined by the ratio between flight size and RTT. It changes with *cwnd* and with the queuing-delay component of the RTT. TCP flavors differ in the way they control *cwnd*, but they all share the general principles of cautious probing for more bandwidth (additive increase) and steep contraction in response to congestion signals (multiplicative decrease). At the congested link, TCP window oscillations induce queue length oscillations, which modulate the queuing delay and the RTT. The bit rate of the TCP connection matches the link capacity $C$ as long as the variations of *cwnd* and RTT compensate each other.

The ratio $\beta$ between the *cwnd* values after and before a multiplicative decrease is of particular interest to buffer sizing. For instance, TCP Reno [14] reduces *cwnd* by 50% ($\beta = 0.5$) and TCP CUBIC [15] drops it by 30% ($\beta = 0.7$). In order for the congested link to remain fully utilized, the decreased *cwnd* must retain a positive queuing delay on top of the propagation component $RTT_0$ of the round-trip time:

$$(RTT_0 + \delta^-) \cdot \beta \geq RTT_0, \qquad (1)$$

where $\delta^-$ is the queuing delay right before decreasing *cwnd* and $Q^- = C \cdot \delta^-$ is the corresponding queue length. Equation (1) yields the following expression for the minimum queue length before the *cwnd* reduction, and therefore for the minimum buffer size $B_{min}$ that guarantees full throughput:

$$B_{min} = Q_{min}^- = C \cdot RTT_0 \cdot \frac{1-\beta}{\beta}. \qquad (2)$$

Equation (2) generalizes the bandwidth-delay product (BDP) rule [16] for a generic TCP flavor with multiplicative decrease ratio $\beta$. The rule yields $B_{min} = C \cdot RTT_0$ with TCP Reno and $B_{min} = 0.4 \cdot C \cdot RTT_0$ with TCP CUBIC. Full utilization of the link capacity is not possible when $B < B_{min}$.

### B. Global Synchronization with Multiple Flows

When $N$ TCP flows share a common bottleneck link the queue length is in equilibrium with the cumulative effect of the $N$ congestion windows. Every congestion signal affects only one flow, causing only one *cwnd* to contract. The resulting drop in bottleneck queue length reflects the current bandwidth-delay product of the affected flow, which becomes smaller as $N$ grows larger. If *cwnd* was guaranteed to be the same for all flows, and congestion signals were spaced in time so that one flow receives one signal not before the queue length has recovered from the previous one, the buffer size could shrink down to $B_{min}/N$. Unfortunately, this is not possible with tail-drop queues because these queues concentrate packet losses for multiple flows within a very short time, causing the contractions of a large portion of *cwnd* instances to overlap, which is exactly what we call global synchronization.

In theory, somewhere in between the single-flow BDP rule and the linear reduction by the number of flows, mildly compressed sizes could be considered safe for tail-drop buffers in high-speed links with large numbers of TCP flows [6], [7]. However, the number of long-lived flows in a link may vary widely in practical scenarios, leaving negligible margins for downsizing a tail-drop buffer that aims at consistently high levels of link utilization during congestion episodes. To achieve more meaningful reductions of queue length and delay, a buffer for high-speed links should disrupt the global synchronization pattern.

We show the basic elements of global synchronization using the example of Fig. 1. All flows simultaneously probe the link for extra bandwidth by gradually increasing their *cwnd*. When the aggregate bit rate of the flows saturates the link capacity, the link enters congestion and the queue size and delay start growing. Any further *cwnd* increase has no effect on the link throughput and only contributes to queue length and delay accumulation. For every TCP Reno flow the queue grows at the rate of one packet per RTT (corresponding to a unit increment of *cwnd*), so with $N$ flows the growth rate is $N$ packets per RTT. With TCP CUBIC flows the growth rate is never lower than with Reno and is frequently higher. We note that in most practical cases the widespread use of delayed acknowledgments by TCP clients [17] and the default host configuration *not to use* the appropriate byte count (ABC) option in TCP senders [18] actually halve the growth rate, e.g., down to $N/2$ packets per RTT with TCP Reno [12].

When the tail-drop queue drops the first packet in a congestion episode, the queue length immediately drops by only one unit. It takes at least an entire RTT before the *cwnd* reduction induced by the packet loss shows its full impact on the queue length. (The time between the first drop and the larger queue length contraction may actually grow close to two RTTs, due to the sub-RTT burstiness of bandwidth-sharing flows [12].)

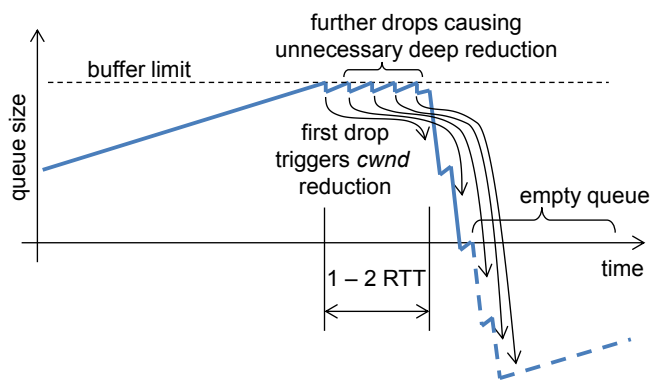

Fig. 1. Synchronization of tail-drop events.

During the RTT interval that follows the first drop event the TCP senders of all flows keep probing for bandwidth at the same pace as before. That is, the packets arriving to the queue exceed those departing by $N/2$ units. Since the queue is already full, it drops $N/2$ packets. If each dropped packet belongs to a different flow, every other flow ends up contracting its *cwnd* at the same time. If the buffer is far smaller than required by the BDP rule, the queue depletes and the link operates at sub-capacity levels until the combined *cwnd* of all flows returns large enough to establish again a continuous presence of packets in the queue. The queue collapse may be less severe when losses hit one or more flows multiple times, so that the fraction of the total population affected by losses is smaller than 50%, but statistically it still presents a problem.

### III. GLOBAL SYNCHRONIZATION PROTECTION

In this section we present three versions of the GSP algorithm: basic, adaptive, and delay-based.

#### A. Basic GSP

Global synchronization can be averted by removing the extra packet drops right after the first one (see Fig. 2). To do so we shift the drop threshold well below the buffer size limit. The first packet drop starts a time interval during which all threshold violations by incoming packets are ignored. Ideally the duration of the interval should be twice as large as the RTT of the dominant flows in the queue (i.e., the flows that contribute the majority of the traffic). The queue is then allowed to keep growing until it feels the effect of the *cwnd* reduction. At the end of the no-drop interval the queue length is well below the drop threshold and requires no further action.

The pseudo-code of Fig. 3 describes the algorithm of Fig. 2. The function `now()` returns the current time. The value of the parameter `interval` is ideally two times the RTT of the traffic that is expected to dominate the queue. The variable `expiry` holds a time value and does not involve the use of a timer.

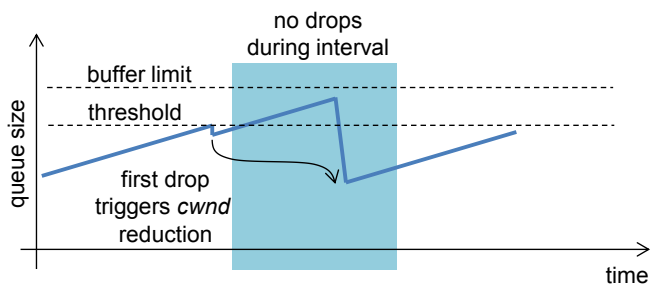

Fig. 2. Global synchronization protection, basic version.

```
at every packet arrival DO:

IF (queue > threshold) AND (now() > expiry)
{
        drop the packet
        expiry = now() + interval
}
ELSE {
        enqueue the packet
}
END
```

Fig. 3. Pseudo-code of the basic GSP algorithm.

A fixed value of `interval` suits well the algorithm when the number of flows in the queue is relatively small, because the queue length contraction after a packet loss is fast and the subsequent recovery is slow. Instead, with many flows and particularly with more aggressive TCP flavors like CUBIC, the queue length may grow faster than it drops after a single loss. When this happens the queue is longer when the no-drop interval ends than it is when the interval starts, so the buffer inevitably overflows and global synchronization returns.

A shorter no-drop interval that expires multiple times per RTT can keep the buffer from overflowing. The queue drops packets periodically as long as the queue length remains above the drop threshold, then stops when the queue shortens, then starts again the periodic drops the next time it crosses the threshold. In this mode of operation the basic GSP algorithm behaves like an on-off (or "bang-bang") controller. The operation is robust against mild deviations from the optimum `interval` setting that anchors the average queue length to the drop threshold. Only larger deviations from the optimum value become disruptive, when the queue length no longer oscillates narrowly around the drop threshold.

In the past the potential risk of synchronization between periodic arrivals and drops has caused skepticism against periodic dropping. For that reason most AQMs today rely on randomized drops. GSP does not need randomization because the periodic-drop regime appears only with large flow numbers. The interleaving of packets from many different flows, together with their sub-RTT burstiness [12], supplies sufficient randomization to the distribution of packet arrivals.

Moreover, the phase of the drop sequence changes randomly with every bang-bang cycle.

*B. Adaptive GSP*

Since the same GSP configuration must work well under most scenarios of practical interest, the scheme must adapt the **interval** value automatically. We choose an adaptation heuristic based on the time that the queue spends above and below the drop threshold.

In *single-drop* operation the queue is most of the time below the threshold and drops a packet only once in many expirations of the maximum interval duration. No adaptation is necessary. The **interval** value must be reduced as soon as the queue starts spending more time above the threshold than below. Let **presetInt** be the initial and maximum setting for the adaptive **interval** variable, **tau** the time constant for the adaptation loop, and **alpha** the emphasis factor for the time spent above the threshold, such that the reaction to load changes is stronger. As a rule of thumb, **tau** should be comfortably larger than **presetInt** (we set the ratio at 5) and **alpha** should not be much larger than 1 (we choose 2 in all our experiments). The steps for adaptation of the **interval** value are listed in Fig. 4. The pseudo-code shows how the algorithmic overhead versus tail-drop remains minimal. Most importantly, GSP never loads the packet memory interface above the line rate because it never drops packets after storing them (as opposed to CoDel [10]).

```
at every packet arrival DO:

cumulTime += (alpha * time_above_threshold –
    time_below_threshold)

cumulTime = min(maxTime, max(0, cumulTime))

interval = presetInt / (1 + cumulTime / tau)

NEXT proceed with basic GSP algorithm
```

Fig. 4. Pseudo-code of GSP adaptation heuristic.

From a control theory perspective the interval adaptation algorithm implements an integral controller on the packet drop rate of the inner control loop (see Fig. 5). The inner control loop just decides whether or not to drop packets at the rate defined by the **interval** value. TCP and the queue react accordingly and feed the current queue size back to the threshold decision.

The control is stable as long as no other source of packet drops is active. One such source is the buffer overflow event, which can synchronize the TCP flows with deep depletions of queue and link load. After the overflow event the adaptive GSP can easily find that the queue spends most time below the drop threshold and inaccurately conclude that no adaptation is necessary. This effect has been observed before for other AQMs [19] and our experiments have confirmed it for GSP. It typically occurs when many new flows start using the queue around the same time. For its mitigation we suspend the accumulation of **time_below_threshold** right after a buffer overflow and resume it again after the queue has completed the cycle from buffer overflow to empty to above threshold. This kind of hysteresis may look rough, but effectively prevents the **interval** value from growing in response to the arrival of new flows (the value must indeed decrease, to break the buffer overflow-depletion cycle).

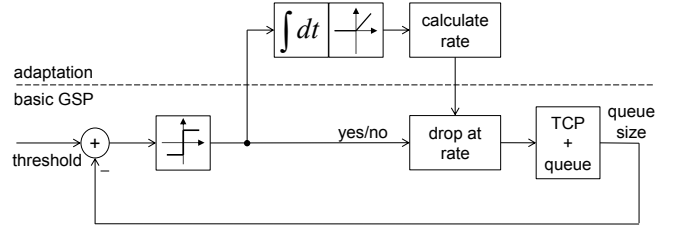

Fig. 5. GSP control architecture.

The adaptation algorithm enables a smooth transition between single-drop and periodic-drop operation. In the periodic-drop regime, the adaptation sets the drop rate based on the queue length placement versus the drop threshold. We underscore that the integral controller that maintains the **interval** value is external to the on-off control loop of the basic GSP algorithm. The internal loop drives the TCP dynamics at the RTT timescale while the external loop modulates one of the internal-loop parameters at a larger timescale. Under steady traffic conditions the outer control loop may very well freeze the **interval** value; instead, the inner loop keeps switching between no-drop and periodic-drop operation, or simply settles on single-drop if **interval = presetInt**.

*C. Delay-Based GSP*

CoDel [10] and PIE [11], AQM schemes of recent introduction, use the queuing delay, not the queue length, as the control target. Could GSP benefit from a similar approach? While the **interval** value controls the stability of the queue by avoiding buffer overflow and global synchronization events, the size of the drop threshold impacts delay statistics and link utilization. If the threshold is too small, even a single packet loss may deplete the queue; if it is too large, a standing queue may form that adds a fixed contribution to the queuing delay of every packet. From Eqs. (1) and (2) we know that the queuing delay budget $\delta^-$ depends only on the RTT, while the minimum queue length $Q_{min}^-$ also depends on the link capacity, thus from a dimensioning perspective it is easier to work with delay than with queue length. Moreover, a delay threshold does not need adjustment when the link capacity changes.

Nevertheless, caution is still required. The physical limit of a buffer is set by the amount of bytes that it can hold. When the link capacity is high, a delay threshold could imply a queue size beyond the buffer size. Just as well, with low capacity a delay threshold could be smaller than the transmission time of a packet. Both cases are dysfunctional.

We enable delay-based operation in GSP by generalizing the meaning of the condition **queue > threshold**. Both terms can be expressed in memory-size units, time units, or a combination of the two. The queuing delay can be measured with one of the methods of CoDel and PIE. CoDel uses timestamps that it associates with packets when they arrive to the queue and then subtracts from the times of departure. PIE estimates the drain rate for translation of the actual queue size into an expected queuing delay. In our experiments we used the timestamp approach.

IV. EXPERIMENTAL EVALUATION

*A. Testbed*

We implemented GSP as a Linux kernel module with both queue-length and delay thresholds and with the adaptation heuristic of Fig. 4. The module enables experimentation in real network conditions and benchmarking against other popular AQM schemes such as CoDel and PIE.

Our evaluation testbed, shown in Fig. 6, consists of four Linux servers (kernel version 3.16) connected by 10GbE links. One server is configured as a router with traffic control enabled on the outgoing interfaces. Within the Linux traffic control subsystem a token-bucket filter serves as a rate limiter, thus creating the bottleneck queue. The queue is controlled by a byte limit in tail-drop experiments and by an AQM plug-in in all other cases (kernel 3.16 versions of CoDel and PIE, and our own GSP module). The end systems implement the RTT emulation and instantiate a configurable number of TCP transmitters (Tx) and receivers (Rx). All TCP flows use CUBIC congestion control, the SACK option, and delayed ACKs. With two servers in parallel we can emulate different RTTs in the same experiment.

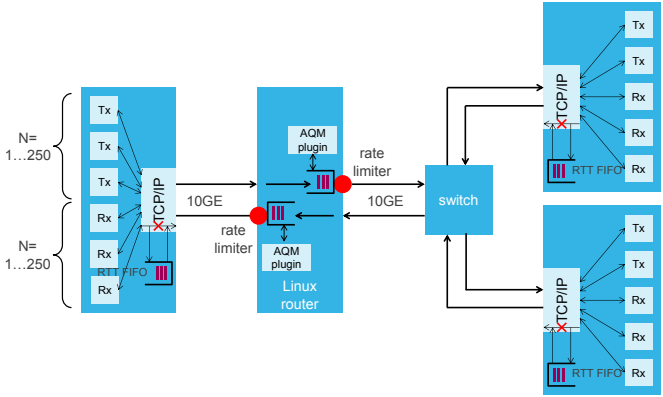

Fig. 6. AQM evaluation testbed.

We capture simultaneous `pcap` traces of packets transiting on both router interfaces. To gain valuable insights into the interplay of TCP traffic and queue management we periodically read out AQM statistics while we inject and monitor test (`ping`) packets.

*B. Queue Operation Examples*

In a first series of experiments we illustrate the operating principles of GSP. We plot queuing delay (computed as the difference between the RTT associated with each returning ACK and the propagation RTT of the data path) and the packet drop events, all extracted from the `pcap` files.

Figure 7 shows a 2s trace from a tail-drop experiment with 10 flows and delay limit smaller than the single-flow delay budget $\delta^-$ from Eq. (1) (12ms versus 40ms for TCP CUBIC). The negative effect of global synchronization is evident. When the queue length saturates the buffer, several packets are dropped before the queue length starts falling. The rate reduction subsequently experienced by multiple flows is deep enough to empty the queue. The link operates at sub-capacity levels for about one second.

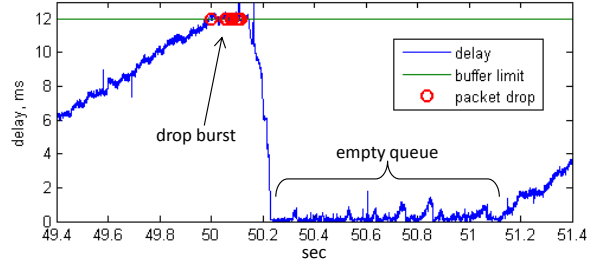

Fig. 7. Tail-drop queue: 10 flows, 100Mb/s link, $RTT_0 = 100$ms.

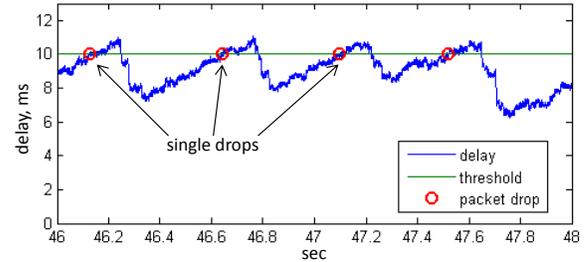

Fig. 8. Basic GSP: 10 flows, 100Mb/s link, $RTT_0 = 100$ms, 10ms threshold.

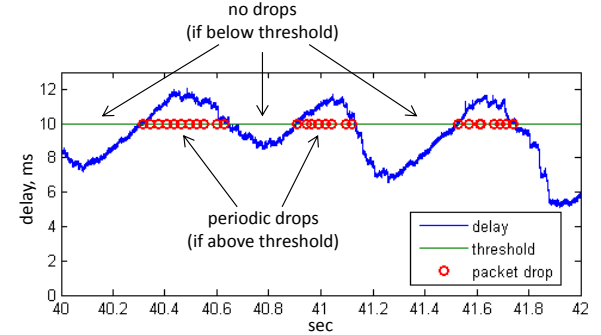

Fig. 9. GSP with interval adaptation: 40 flows, 400Mb/s, $RTT_0 = 100$ms, 10ms threshold.

Figure 8 shows how the basic GSP algorithm avoids the synchronization in the same scenario of Fig. 7. After the first packet drop event the no-drop interval allows the queue to grow further without experiencing new losses. Only one of the ten parallel flows reduces its *cwnd*. The subsequent queue reduction is much smaller than in the synchronized case.

In the experiment of Fig. 9 we increase the number of flows from 10 to 40 (the link capacity also grows, from 100Mb/s to 400Mb/s). The plot shows that the growth rate of the queue length is now too large for the 200ms `interval` value of the basic GSP to keep the queue in a stable equilibrium. The adaptation algorithm of Section III.B becomes necessary to increase the frequency of the packet-drop decisions, so that several losses occur before the queue length starts falling. The queue alternates between periodic-drop and no-drop periods. The on-off control holds the queue in equilibrium while the adaptation algorithm slowly adjusts the drop rate.

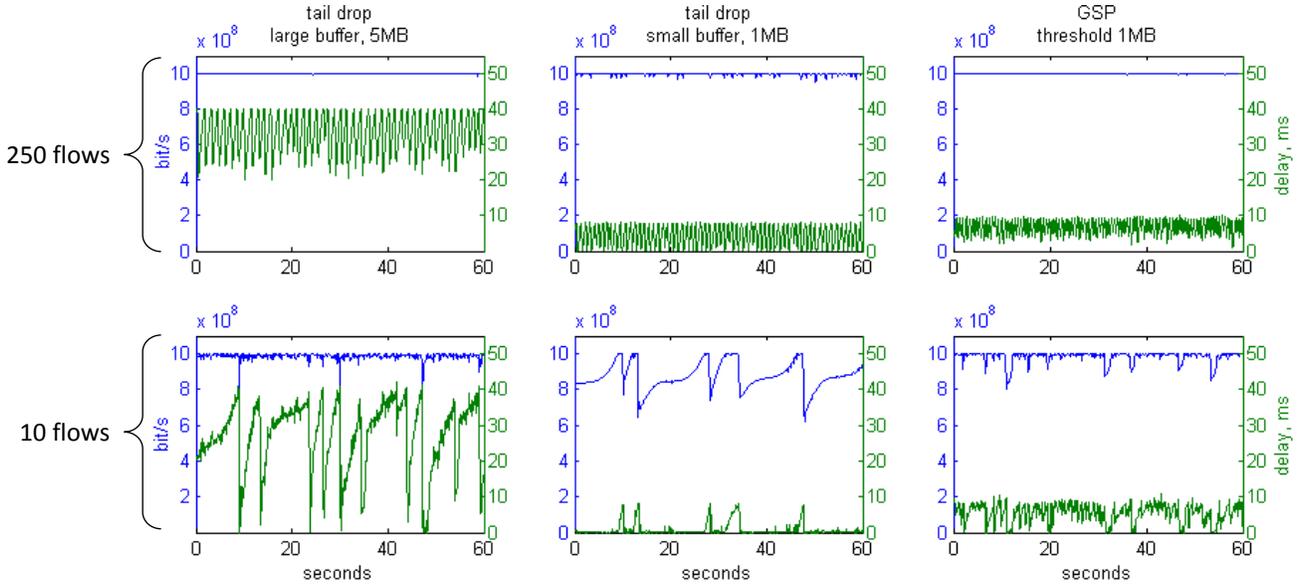

Fig. 10. GSP compared to large and small tail-drop buffers at high and low flow numbers.

### C. Performance with Different Flow Numbers

In this set of experiments we show that tail-drop works well in small buffers if the number of flows is large [6][7], but fails to fully utilize the link when it is loaded with fewer flows. All experiments use a 1Gb/s link with $RTT_0 = 100\,\text{ms}$.

The plot on the top left of Fig. 10 is obtained with a 'large' buffer, sized for TCP CUBIC according to Eq. (2): $B_{min} = 5\,\text{MB}$. The link is always loaded to its full capacity, but the buffer is clearly too large for 250 flows, so all packets experience unnecessary extra delay (at least 20ms). In contrast, with only 10 flows in the mix (bottom left) the onset of global synchronization causes the queue length to oscillate over the entire range (0-40ms). The plots in the center column are from the same tail-drop setup, except for the buffer size, now set to 1MB, or 20% of the $B_{min}$ value for CUBIC. With 250 flows (top) the queue keeps the link fully loaded. The queuing delay oscillates below 8ms. With only 10 flows (bottom), the link utilization drops to a minimum of 60% and an average of 87%. The small buffer is empty most of the time. With the drop threshold also set to 1MB, GSP (right column) keeps the queuing delay always below 10ms irrespective of the number of flows. The average link utilization is 100% with 250 flows (top right of Fig. 10) and 98% with 10 flows (bottom right).

### D. Steady-State Performance

In this section we compare GSP with CoDel [10] and PIE [11]. We focus on the AQM's ability to keep the queuing delay low around a target value without losing throughput when the queue depletes. In Fig. 11 we show the probability distributions of the queuing delay under different multiplexing degrees (1, 10, and 100 flows, always with a per-flow average fair share of 10Mb/s). For GSP, CoDel, and PIE we set the buffer size to the bandwidth-delay product (125kB, 1.25MB, and 12.5MB). For tail-drop we set the size to a CUBIC-optimized value of 40% of BDP (50kB, 500kB, 5MB). CoDel and PIE use their default Linux values for all other parameters.

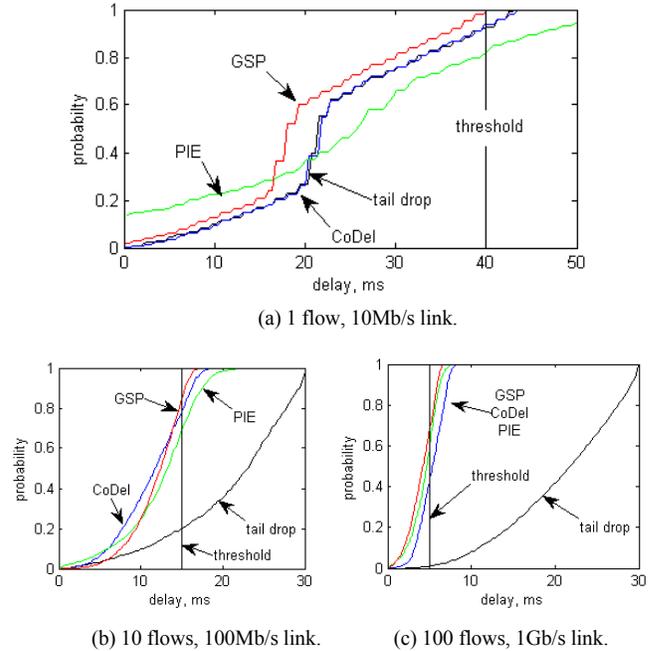

Fig. 11. CDF of queuing delay for 1, 10, and 100 flows; $RTT_0 = 100$ms.

For best AQM operation under congestion, the cumulative density function (CDF) should start at zero, indicating that the queue never depletes. Then, as the delay increases, the CDF should reach probability one as steeply as possible, meaning that the delay remains low under all circumstances. Figure 11(a) confirms that no AQM can do better than a well-sized tail-drop queue when only one flow is present: the throughput is lower and the delay distribution is not better despite the smaller traffic volume (see in particular the PIE curve). Figures 11(b) and 11(c) show that all AQMs improve the delay distribution as the number of flows increases. In both cases GSP fares really well compared to CoDel and PIE.

The results of Fig. 11 are remarkable because GSP is designed exclusively around the goal of suppressing global synchronization. It was well expected that GSP could not do better than tail-drop in the single-flow case. In the plain multi-flow scenarios of the experiment, GSP always performs at least as well as CoDel and PIE, and even better in some cases.

*E. RTT robustness*

In Section III.A we indicated that the basic GSP should run with **interval** set to twice the dominant RTT. In the adaptive version of GSP the same value should be chosen for **presetInt**, which is the initial and maximum value of **interval**. In practice the choice of the value is a matter of coarse approximation. A smaller-than-expected dominant RTT causes faster queue length oscillations, calling for a smaller value of **interval** that the adaptation promptly provides. Setting **presetInt** at 200ms should work well in all cases where a dominant RTT smaller than 100ms is not guaranteed to be enforced.

To test the robustness of CoDel, PIE, and GSP against RTT variations we keep fixed configuration parameters while changing the RTT for a set of 10 flows in a 100Mb/s link. The values of all parameters are the default ones, except for the drop threshold, which we set at 5ms for all schemes. Fig. 12 shows that with RTT at 10, 20, and 50ms GSP converges to the 5ms drop threshold equally or even better than the other schemes. With 100ms RTT the aggressive drop threshold causes a slight throughput reduction (to 99%) for all schemes, and therefore a null value of the 5% delay quantile.

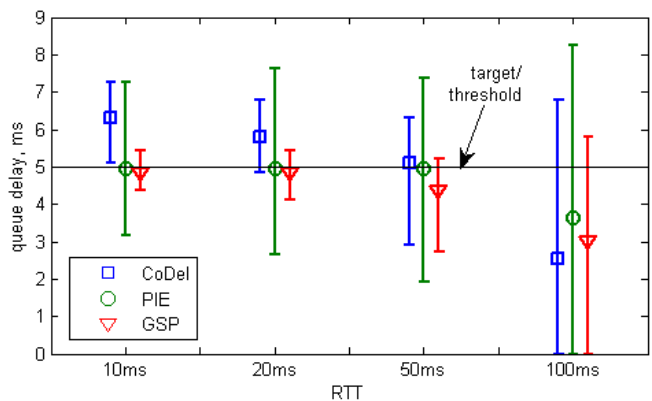

Fig. 12. Adaptation to deviating RTT; error bars show median, 5% and 95% quantiles of the queue delay.

We have also run experiments with mixed RTT values in the same queue (10ms and 100ms). The results (not shown here for lack of space) confirm the well-known RTT bias of TCP for all single-queue approaches, without remarkable differences between GSP and CoDel/PIE.

*F. Variable Transmission Capacity*

A desirable feature of delay-based AQMs is the ability to adapt to undefined or variable link capacities. In the next experiment we look at the queue response to a capacity drop from 100Mb/s to 10Mb/s, and then to the reverse transition from 10Mb/s to 100Mb/s. There are 10 flows sharing the bottleneck link, but the buffer size is the single-flow CUBIC optimum at 100Mb/s and 100ms propagation RTT (500kB).

Fig. 13 shows that in the first 30s of the experiment, with link capacity at 100Mb/s, the queuing delay is well confined below 40ms. After the capacity falls to 10Mb/s, both GSP and tail-drop experience a delay spike. The larger delay is measured for packets that are already queued at the time of the transition and for those that arrive before the TCP senders detect the packet loss acceleration: there is not much that an AQM can do to avoid this transient effect besides increasing the frequency of the packet drop decisions. GSP absorbs the transition in about 5s and quickly brings back the queuing delay around the drop threshold (set at 15ms). Instead, since tail-drop anchors the queue length to the buffer size, its delay now oscillates in the 300-400ms range. The queue depletion seen after the initial 100Mb/s capacity is restored is also unavoidable by a buffer that is reasonably sized, as it is entirely controlled by the speed of the *cwnd* recovery at the TCP transmitters. A much larger buffer size or drop threshold could keep the *cwnd* distribution at the value needed to avoid the buffer depletion, but would also induce unbearable delays when the link capacity drops. This approach is considered acceptable across wireless links, where capacity variations are continuous and the number of competing flows is small, but would be overly detrimental in high-speed core links.

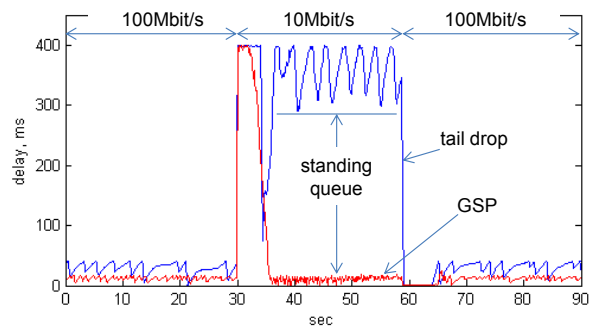

Fig. 13. Queuing delay at variable link capacity with tail-drop and delay based GSP.

*G. Unresponsive Traffic*

AQM algorithms assume that all traffic in the buffer responds to congestion signals, so they lose effectiveness when a fraction of the traffic does not respond as expected. Different schemes may not have the same ability to compensate for diversions from the ideal mode of operation.

The experiment of Fig. 14 mixes TCP and UDP traffic in the same queue. We start the experiment with 10 TCP flows loading a 100Mb/s link. After a while we add a 90Mb/s UDP flow from a constant-bit-rate source. The queue saturates at the tail-drop limit right after the UDP traffic starts. This cannot be avoided because of the excess TCP packets that are already in flight. All AQMs eventually return the queue to the target delay level. PIE shows the fastest reaction but also wide oscillations around the new equilibrium. GSP shows the slowest reaction with the narrowest oscillations. The parameter `tau` defines the tradeoff between stability and agility under changing traffic conditions. The stability favored by the setting used in our experiments (`tau` = 5 × `presetInt` = 1s) is well justified in a high-speed link, where traffic mix variations are typically not as steep as the one applied in this experiment.

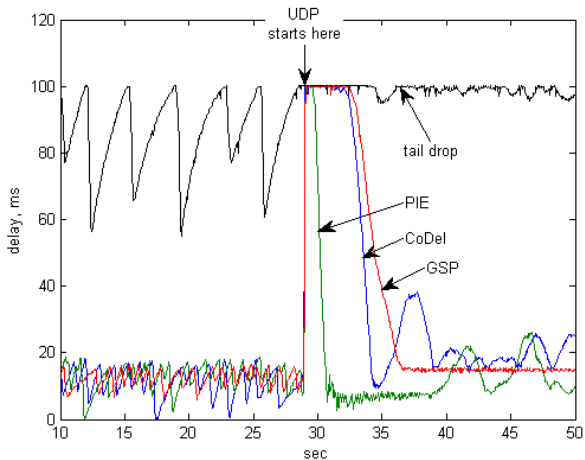

Fig. 14. Queue reaction to a sudden UDP injection (90% of link capacity).

## V. CONCLUSION

We presented a new minimalistic AQM algorithm called global synchronization protection (GSP) that requires only few additional operations in the fast path of a tail-drop packet queue. All processing steps added by GSP, including the packet drop decision, occur during the packet-enqueue phase. By not loading the buffer interface above the line rate, GSP proves very appealing for packet processors that operate at leading-edge rates. The design of GSP is motivated by the well-known phenomenon of synchronization among the congestion window cycles of TCP flows that share a tail-drop buffer. The phenomenon causes large queue length oscillations and adds disruptive queuing delays to the cost of throughput preservation. GSP safely breaks the synchronization when the number of flows is small. With more flows it smoothly transitions into an on-off control that keeps the queue length within the range of a preset target. The transition is driven by the adaptation of a single parameter in the slow path. GSP can work with both queue-length and queuing-delay thresholds. The latter mode of operation is advantageous when the drain rate of the queue is unknown or simply variable.

We implemented a Linux kernel module as a proof-of-concept prototype and performed numerous experiments in a testbed with real network equipment. The experiments expose the operating regimes of GSP and favorably report on the performance of the new scheme when compared to other single-queue AQMs of recent introduction, namely CoDel and PIE. They also highlight the benefits of delay-based operation at variable link rates and the ability to isolate TCP traffic from unresponsive flows.

We are now exploring further enhancements of the adaptation heuristic and ways to automatically adapt the drop threshold to the traffic mix. We are also looking at more complex traffic scenarios where congestion is present in both directions of the data path.

ACKNOWLEDGMENT

This work has been funded in part by the German Bundesministerium für Bildung und Forschung (Federal Ministry of Education and Research) in scope of project SASER under grant No. 16BP12200.